\documentclass[%
 amsmath,amssymb,
 aps, physrev,
]{revtex4-2}

\usepackage{graphicx}
\usepackage{dcolumn}
\usepackage{bm}
\usepackage{graphicx} 
\usepackage{xcolor}
\usepackage[utf8x]{inputenc}
\usepackage{physics}
\usepackage{siunitx}
\usepackage{textgreek}
\usepackage[version=4]{mhchem}
\usepackage{comment}
\usepackage{lipsum}
\usepackage{upgreek}
\usepackage{float}

\begin{document}

\title{\textbf{Translational dynamics of lipid-coated microbubbles driven by ultrasound}}

\author{Marco Cattaneo}
 \email{Contact author: mcattaneo@ethz.ch}
\author{Outi Supponen}%
\affiliation{Institute of Fluid Dynamics, ETH Zürich, Zürich, Switzerland}

\begin{abstract}
Ultrasound-driven microbubbles are increasingly being investigated for both molecular imaging and therapeutic applications. 
To be effective, these bubbles must be brought into close proximity or direct contact with the target site. 
Leveraging the acoustic radiation force provides a powerful strategy to direct their movement.
In this study, we examine the translational dynamics of a microbubble with unprecedented accuracy by simultaneously time-resolving both its radial and translational responses and by using optical tweezers to study the bubble in free space.
Our experimental results show excellent agreement with theoretical predictions for the bubble sizes considered, provided the history drag force is included in the force balance. 
For the Reynolds numbers considered (up to $\text{Re}\approx2$), the zero-Reynolds-number history force performs as well as its finite-Reynolds-number extension.
Although non-spherical modes may arise at larger bubble expansions, they do not appear to significantly influence the bubble translational motion.
A major finding is that the normalised transport distance of the bubble scales linearly with the normalised volumetric expansion during its oscillation, greatly simplifying the design and analysis of transport strategies.
We also investigated bubble stability during transport and observed a marked increase in dissolution rate once a threshold in bubble expansion is exceeded. 
These insights can be leveraged to develop optimal transport strategies that balance both transport speed and bubble stability for targeted delivery applications.
\end{abstract}

\maketitle

\section{Introduction}

Ultrasound contrast agent microbubbles have transformed medical imaging by enhancing blood echogenicity, significantly improving the visualisation of vasculature and perfusion in clinical diagnostics.
These microscopic gas-filled spheres are coated by a stabilising shell, typically composed of phospholipids, to reduce gas efflux and extend circulation time \cite{Ferrara2007UltrasoundDelivery}.
Beyond contrast-enhanced imaging, coated microbubbles are increasingly explored for molecular imaging and therapeutic applications.
Functionalising their surface with targeting ligands enables selective binding to specific tissues or pathological markers, facilitating precise localisation and potentially enabling targeted drug delivery \cite{Shakya2024Ultrasound-responsiveDelivery}.
Additionally, microbubbles can serve as cavitation nuclei in sonothrombolysis, aiding in blood clot breakdown under high-intensity ultrasound \cite{ElKadi2022ContrastDisease,Bautista2023CurrentSonothrombolysis}.
However, for these applications to be effective, microbubbles must come into close proximity or direct contact with the target site \cite{Ward2000ExperimentalVitro,Zhou2012ControlledCavitation}.
Yet, most systemically circulating microbubbles flow within the blood vessel lumen, distant from where therapeutic action is needed.
To address this, the primary acoustic radiation force, or primary Bjerknes force, has been employed to direct microbubble movement \cite{Shortencarier2004ALipospheres,Rychak2005AcousticVerification}.
This force arises from spatial gradients in the ultrasound pressure field, resulting in a net force that propels the microbubbles along the direction of ultrasound propagation \cite{Prosperetti1982BubbleResults}.
However, the successful implementation of this approach is not trivial.
Coated microbubbles are delicate structures prone to dissolution under prolonged or intense ultrasound exposure.
During the compression phase of the oscillation, pressure in the bubble gas core rises, driving gas out of the bubble. 
In addition, when the relative oscillation amplitude exceeds approximately 0.3, some lipids may shed off, reducing the lipid-coated surface area \cite{Luan2014}.
Maintaining microbubble integrity during their displacement is essential, as their size relative to the driving ultrasound frequency determines their mechanical effectiveness.
Thus, developing strategies to efficiently transport microbubbles while preserving their stability is critical for biomedical applications.
A key challenge in achieving this is the limited availability of time-resolved data capturing both the translational and radial dynamics of ultrasound-driven microbubbles in a free space, which are needed for validating coupled theoretical models and, ultimately, for optimising transport to achieve a desired displacement while keeping radial expansion within stability limits to reduce dissolution.

Dayton \textit{et al.} \cite{Dayton2002TheAgents} pioneered the use of streak camera imaging to investigate the radial and translational dynamics of single phospholipid-coated microbubbles.
In their experiments, bubbles travelled along the wall of cellulose tubing, and the authors compellingly demonstrated that a single microbubble could be displaced by more than \SI{5}{\micro\meter} at speeds exceeding \SI{0.5}{\meter\per\second} under the influence of a 20-cycle, \SI{380}{\kilo\pascal}, \SI{2.25}{\mega\hertz} ultrasound pulse.
The measured displacements were overestimated by approximately 20–30\% in their simulations, a discrepancy likely attributable to wall friction or additional unaccounted forces—highlighting the need for refinement in the physical modelling.
Building on these insights, Vos \textit{et al.} \cite{Vos2007MethodForce} employed ultrasound imaging to track individual microbubble displacements across a wide range of acoustic pressures (\SI{150}{\kilo\pascal} to \SI{1.5}{\mega\pascal}) and frequencies (2–8 MHz), offering valuable information on the mechanical response of the microbubble shell.
Although this approach did not capture the equilibrium radius or radial oscillations, it significantly advanced the understanding of translational behaviour.
A similar methodology was adopted by Supponen \textit{et al.} \cite{Supponen2020ThePopulations}, with the added benefit of prior size-isolation, enabling more accurate estimation of the bubble radius.
Meanwhile, Segers and Versluis \cite{Segers2014AcousticEnrichment} explored the application of the primary radiation force to selectively isolate narrow size distributions from polydisperse microbubble populations in microfluidic channels.
Their experiments utilised low-pressure (\SI{10}{\kilo\pascal}) continuous ultrasound, specifically focusing on time-resolving the displacement dynamics.
Acconcia \textit{et al.} \cite{Acconcia2018TranslationalPulses} introduced a novel approach by using optical tweezers to position single bubbles away from nearby surfaces, enabling the study of translational motion in free space under millisecond-long ultrasound pulses (1 MHz, 25–200 kPa).
Although radial dynamics were not explored experimentally in this work, the setup offers a promising framework for investigating unconstrained bubble motion.
Finally, Garbin \textit{et al.} \cite{Garbin2009HistoryUltrasound} made significant contributions by employing ultra-high-speed imaging to resolve both radial and translational bubble dynamics, but with an emphasis on exploring the secondary radiation force between bubble pairs under acoustic excitation.

The theoretical models describing the translation of ultrasound-driven bubbles combine an equation for radial dynamics with another one that accounts for the balance of forces along the direction of translation.
The radial dynamics equation is based on the Rayleigh--Plesset equation \cite{Rayleigh1917Cavity,Plesset1949TheBubbles} or its variants \cite{Brenner2002Single-bubbleSonoluminescence}, modified to account for the presence of the shell, which is often achieved using the Marmottant nonlinear model \cite{Marmottant2005ARupture}.
The force balance equation includes contributions from the bubble inertia, acoustic forcing, and hydrodynamic forces.
The latter consist of the added-mass force, which arises from the instantaneous acceleration of the displaced fluid, the quasi-steady drag force, dependent solely on the instantaneous velocity of the bubble, and the history drag force, which results from the unsteady diffusion of vorticity from the bubble surface.
This force is named so because, at any given moment, it depends on the entire history of the bubble translational and radial accelerations.
It becomes particularly significant at low Reynolds numbers.
For a bubble with a fixed radius, Yang and Leal \cite{Yang1991ANumber} derived an analytical expression for the history force under the assumption of unsteady creeping motion (zero Reynolds number). 
Magnaudet and Legendre \cite{Magnaudet1998TheRadius} later generalised this result to bubbles with varying radii and shear-free interfaces by applying a transformation of space and time coordinates. 
Their analysis revealed that, when the bubble radius changes over time, a history force persists even in the absence of relative acceleration between the bubble centre and the surrounding fluid. 
Takemura and Magnaudet \cite{Takemura2004TheNumber} extended this work further by deriving an expression for the history force for bubbles with varying radii that obey a no-slip condition, such as surfactant-covered bubbles.
They also provided a formulation for the history force at finite Reynolds numbers.

However, a comprehensive comparison of the various models for the history force has not yet been performed for acoustically driven microbubbles, nor has the validity range been clearly defined in terms of Reynolds numbers.
Furthermore, when the oscillation amplitude exceeds a certain threshold, bubbles begin to exhibit non-spherical deformations (shape modes) \cite{Cattaneo2025CyclicDelivery,Cattaneo2025ShapeBubbles} due to the Faraday instability \cite{Faraday1831XVII.Surfaces}, in addition to their volumetric oscillations (breathing mode).
The influence of these shape modes on bubble displacement remains unexplored.
Finally, perhaps the most crucial aspect of bubble transport that has yet to be investigated is the extent of bubble dissolution induced by ultrasound driving, which will determine the final bubble size at the target location and, consequently, its efficacy, as a significantly smaller bubble will be out of resonance with the driving frequency and exhibit much weaker oscillations.
In this study, we address these gaps by simultaneously resolving the translation and oscillation of single microbubbles under microsecond-long ultrasound pulses using ultra-high-speed microscopy imaging.
Optical tweezers are used to isolate the microbubbles from boundaries, allowing us to study their behaviour in free space.
The primary objectives are:  
(i) to compare experimental data with theoretical predictions to determine which models perform best,  
(ii) to assess the effect of shape modes on the overall translational motion, 
(iii) to derive a simple scaling law that links radial behaviour with translational motion, and  
(iv) to evaluate bubble stability as a function of transport speed, aiming to develop an optimal transport strategy that balances speed and bubble integrity.

\section{Experimental setup}

\begin{figure}[t]
    \includegraphics[width = 0.5\columnwidth]{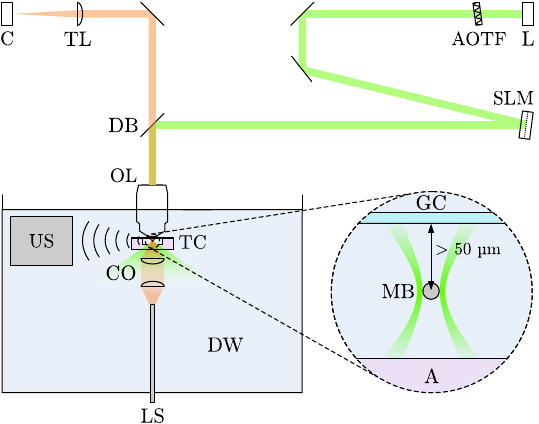}
     \caption{Experimental setup. (A) Agarose, (AOTF) Acousto-optic tunable filter, (C) Camera, (CO) Condenser, (DB) Dichroic beamsplitter, (DW) Deionised water, (GC) Glass coverslip, (L) Laser, (LS) Light source, (MB) Microbubble, (OL) Objective lens, (SLM) Spatial light modulator, (TC) Test chamber, (TL) Tube lens, (US) Ultrasound transducer. 
     The inset provides a close-up view of the pre-test conditions, showing a single optically-trapped microbubble positioned more than \SI{50} {\micro\meter} from the test chamber's coverslip.}
     \label{fig:Figure1}
\end{figure}

The experimental setup is sketched in Fig.~\ref{fig:Figure1}.
Commercially available phospholipid-coated microbubbles (SonoVue, Bracco), with equilibrium radii in the range $R_0=\SIrange{1.8}{4.9}{\micro\meter}$, are introduced into the test chamber at a highly diluted concentration to avoid inter-bubble radiative forces (secondary Bjerknes forces) during acoustic driving.
The test chamber features bottom and side walls made of agarose 1\%, ensuring acoustic transparency due to its acoustic impedance being nearly identical to that of water \(\left( \mathcal{Z} = \SI{1.5e6}{\pascal\second\per\meter} \right)\).
A microscope glass coverslip seals the top of the chamber, serving as the imaging window for the bubbles.
The test chamber is immersed in a deionised water bath (temperature $T_{\mathrm{l}} = \SI{22}{\celsius}$).
Due to buoyancy, the bubbles rise and adhere to the coverslip.
Single bubbles are optically trapped and displaced by more than $\SI{50}{\micro\meter}$ from the coverslip using optical tweezers to minimise acoustic interference from the wall; at this distance ($z>10R_0$), the resonance shift induced by the rigid wall $\left(f_{\rm wall}/f_0 = 1/\sqrt{1+R_0/2z}\right)$ is below $2.5\%$.
The optical tweezers system utilises a 532-nm continuous wave laser (Verdi G10, Coherent) to reduce water heating compared to infrared sources.
A reflective spatial light modulator (SLM) (PLUTO-2.1 VIS-096, HOLOEYE Photonics) shapes the beam into a Laguerre-Gaussian profile, which is necessary for trapping particles with a refractive index lower than the surrounding medium, such as bubbles.
The trap dimensions and laser intensity are manually adjusted to match each bubble size. 
Finally, a dichroic mirror directs the modulated beam into the objective lens, which focuses it onto the sample. 
For a detailed description of the system, please refer to our previous work \cite{Cattaneo2023ShellMicrobubbles}.
The bubble is driven acoustically at a frequency $f_{\rm d} = \SI{1.5}{\mega\hertz}$ using an ultrasound transducer (PA2366, Precision Acoustics)  oriented parallel to the horizontal plane.
The driving pulse is generated by a function generator (LW 420B, Teledyne LeCroy) and subsequently amplified by a radiofrequency power amplifier (1020L, E\&I).
The acoustic driving pressure is measured \emph{in situ} within the test chamber using a calibrated needle hydrophone ($\SI{0.2}{\milli\meter}$, NH0200, Precision Acoustics), which is also used to align the acoustic focal point with the optical field.
Any reflections that may arise from a slight tilt of the coverslip are therefore inherently included in the reported pressure amplitude.
Just before initiating ultrasound excitation, the optical trap is deactivated using a microsecond-fast electronic shutter based on an acousto-optic tunable filter (AOTF.NC-VIS/TN, AA Opto Electronic).
The bubble radial and translational response is recorded from a top-view perspective using a custom-built microscopy imaging system.
This system features a water-dipping objective lens (CFI Plan 100XC W, Nikon; \SI{2}{\milli\meter} focal length) and a tube lens (TL400-A, Thorlabs; \SI{400}{\milli\meter} focal length), achieving a total magnification of $200\times$. An ultra-high-speed camera (HPV-X2, Shimadzu) captures images at 10 million frames per second over $\SI{25.6}{\micro\second}$, with a pixel resolution of 160 nm.
For live imaging, backlight illumination is supplied by a continuous halogen illuminator (OSL2, Thorlabs), while sequentially operated Xenon flash lamps (MVS-7010, EG\&G) provide illumination for video recording.
These light sources are combined into a single optical fibre output and directed onto the sample through a custom-built condenser.
The ultrasound pulse activation, camera recording, light flash, and optical trap deactivation are synchronised using a delay generator (DG645, Stanford Research Systems).
To ensure an accurate comparison between experimental results and theoretical predictions, all bubble radius measurements in this study have been corrected by subtracting \( R_{\mathrm{err}} = \SI{140}{\nano\meter} \). 
This adjustment is based on our previous finding that bright-field microscopy systematically overestimates the radius due to Fresnel diffraction effects.
The correction value was determined using fluorescence microscopy measurements and is specific to the optical setup employed \cite{Cattaneo2023ShellMicrobubbles}.

\section{Theory}

The radial dynamics of a spherical bubble with a time-dependent radius \( R(t) \) is described by a modified Rayleigh–Plesset equation, adapted for mildly compressible Newtonian media \cite{Prosperetti1986BubbleTheory,Lezzi1987BubbleTheory,Brenner2002Single-bubbleSonoluminescence,Prosperetti2013AOscillations}. 
This formulation incorporates an additional pressure term, \( \Sigma\bigl(R, \dot{R} \bigr) \), which accounts for generalised interfacial stresses, and is given by:
\begin{equation}\label{eq:RP}
\rho_{\mathrm{l}} \left(R \ddot R + \frac{3}{2} \dot R^2 \right)= \\ \left( 1 + \frac{R}{c_{\mathrm{l}}} \frac{d}{dt} \right) p_g
   +  \Sigma\bigl(R, \dot R\bigr) - p_{\infty} - p_{\mathrm{d}}\left(t\right) - 4\mu_{\mathrm{l}} \frac{\dot R}{R},
\end{equation}
where over-dots represent time derivatives, \( \rho_{\mathrm{l}} = \SI{997.8} {\kilo\gram\per\cubic\meter}\) is the liquid density, \( c_{\mathrm{l}} = \SI{1481} {\meter\per\second}\) is the speed of sound in the medium and, \( \mu_{\mathrm{l}} =\SI{9.54e-4} {\pascal\second}\) is the dynamic viscosity of the medium.
The gas pressure inside the bubble is given by \( p_{\mathrm{g}} \), while \( p_{\infty} = \SI{102.2} {\kilo\pascal} \) represents the undisturbed ambient pressure.
The term \( p_{\mathrm{d}}(t) \) corresponds to the ultrasound driving pressure.
Due to the ultrasound wavelength $\lambda$ being significantly larger than the bubble size ($\lambda/2R_0>100$), the driving pressure can be considered spatially uniform concerning radial dynamics.
Additionally, since the acoustic wave propagates much faster than the bubble translation velocity $\dot{s}$ ($c_{\mathrm{l}}/\dot{s}>1000$), the dependence of the driving pressure on the bubble position can be neglected.
The pressure contribution from the phospholipid-coated interface is modelled using the Marmottant model \cite{Marmottant2005ARupture}:  
\begin{equation}
 \Sigma\bigl(R, \dot R\bigr) = - 2\frac{\sigma (R)}{R} -4\kappa_{\mathrm{s}}\frac{\dot R}{R^2}, \quad
     \text{where}  \ \sigma\bigl(R\bigr) = 
\begin{cases}
 0, & \text{if } R\leq R_{\mathrm{buckling}},\\[1mm]
 \sigma_0 + E_{\mathrm{s}} \left(J-1\right), & \text{if } R_{\mathrm{buckling}}< R < R_{\mathrm{rupture}},\\[1mm]
 \sigma_{\mathrm{water}}, & \text{if } R \geq R_{\mathrm{rupture}}.
\end{cases}
\end{equation}
Here, \(\sigma(R)\) represents the interfacial tension, while \(\kappa_{\mathrm{s}}\) denotes the interfacial dilatational viscosity.
The surface tension is defined piecewise: it is zero for radii below the buckling threshold, matches the surface tension of a clean gas-water interface $\sigma_{\mathrm{water}} = \SI{72.8}{\milli\newton\per\meter}$ beyond the rupture radius, and varies linearly with the relative area deformation \(J = R^2 / R_0^2\) in the intermediate regime, with a slope determined by the interfacial dilatational modulus, \(E_{\mathrm{s}}\) and a surface tension at the equilibrium radius, $\sigma_0$. 
The internal bubble pressure, \( p_{\mathrm{g}} \), is modelled using the Zhou model \cite{Zhou2021ModelingBubble}, as the commonly used polytropic approximation neglects thermal damping and becomes inaccurate when the Péclet number approaches one—conditions that apply to microbubbles driven by ultrasound at megahertz frequencies \cite{Prosperetti1988NonlinearDynamics}.
This method assumes that the gas pressure inside the bubble remains uniform \citep{Prosperetti1988NonlinearDynamics}.
For an ideal gas, the radial velocity \( u_{\mathrm{g}}(r) \) is given by:  
\begin{equation}
u_{\mathrm{g}}(r) = \frac{1}{\gamma p_{\mathrm{g}}}\left(\left(\gamma-1\right) K_{\mathrm{g}} \frac{\partial T_{\mathrm{g}}}{\partial r} -\frac{1}{3} r \dot{p}_{\mathrm{g}} \right),
\end{equation}
which leads to an exact expression for the gas pressure:
\begin{equation}
\dot{p}_{\mathrm{g}} = \frac{3}{R}\left(\left(\gamma-1\right) K_{\mathrm{g}} \left.\frac{\partial T_{\mathrm{g}}} {\partial r}\right|_{R} -\gamma p_{\mathrm{g}} \dot{R} \right),
\end{equation}
where \( r \) is the radial coordinate, \( \gamma \) is the gas specific heat ratio, \( K_{\mathrm{g}} \) is the thermal conductivity of the gas, and \( T_{\mathrm{g}} \) is the gas temperature.  
The temperature distribution inside the bubble is considered divided into three regions: a core region with uniform temperature, a buffer layer, and an outer region with a linear temperature gradient.
Since the temperature change at the bubble surface is negligible, it is approximated as \( T_{\mathrm{g}}|_{R} \approx T_{\mathrm{l}} \) \citep{Prosperetti1988NonlinearDynamics}.
The volume-averaged temperature \( T_{\mathrm{g}_{\textit{i}}} \) in each region \( i \) is determined using the ideal gas law:
\begin{equation}
T_{\mathrm{g}_{\textit{i}}} = \frac{p_{\mathrm{g}}}{\rho_{\mathrm{g}_{\textit{i}}}\mathcal{R}}, \quad \text{for } i = 1,2,3,
\end{equation}
where \( \rho_{\mathrm{g}_{\textit{i}}} \) is the volume-averaged gas density in region \( i \), and \( \mathcal{R} \) is the specific gas constant.  
The gas density \( \rho_{\mathrm{g}_{\textit{i}}} \) is computed using the continuity equation for each region:
\begin{equation}
\dot{m}_{\mathrm{g}_1} = -f_1, \quad \dot{m}_{\mathrm{g}_2} = f_1 - f_2, \quad \dot{m}_{\mathrm{g}_3} = f_2,
\end{equation}
where \( m_{\mathrm{g}_{\textit{i}}} \) represents the gas mass in region \( i \), and \( f_{j} \) is the mass flux across interface \( j \), given by:
\begin{equation}
f_j = \rho_{\mathrm{g},\textit{j}}^{\mathrm{uw}} u_{\mathrm{g},\textit{j}}^{\mathrm{rel}} S_j,  \quad \text{for } j = 1,2,
\end{equation}
where \( \rho_{\mathrm{g},\textit{j}}^{\mathrm{uw}} \) is the density on the upwind side of interface \( j \), \( u_{\mathrm{g},\textit{j}}^{\mathrm{rel}} \) is the convective velocity (the difference between the real gas velocity and the interface velocity), and \( S_j \) is the interface surface area.  
Due to the extended residence time (approximately ten minutes) of the microbubbles in an air-saturated liquid, their original gas core is gradually replaced by air \cite{Kwan2012LipidExchange}.
Therefore, the gas properties are set to \( \gamma = 1.4 \) for the specific heat ratio, \( K_{\mathrm{g}} = \SI{0.026} {\watt\per\meter\per\kelvin} \) for thermal conductivity, and \( \mathcal{R} = \SI{287} {\joule\per\kilo\gram\per\kelvin} \) for the specific gas constant.

The translational straight-line motion $s(t)$ of a spherical microbubble with radius $R(t)$, driven by ultrasound and moving along the direction of ultrasound propagation, is governed by the balance of forces acting on it, including its inertial force $F_{\mathrm{I}}(t)$, the primary acoustic radiation force \( F_{\mathrm{US}}(t) \), the added mass force \( F_{\mathrm{AM}}(t) \), and the viscous drag force \( F_{\mathrm{D}}(t) \):
\begin{align}\label{eq:forcebalance}
{{F}_{\mathrm{I}}(t)} =  {{F}_{\mathrm{US}}(t)} +  {{F}_{\mathrm{AM}}(t) + {F}_{\mathrm{D}}(t)}.
\end{align}
The inertial force of the bubble is given by:
\begin{equation}
{F}_{\mathrm{I}}(t) = m\ddot{{s}}.
\end{equation}
where $m=\rho_{\rm g,0}V_0$ is the bubble mass, $\rho_{\rm g,0}=\SI{1.225}{\kilo\gram\per\cubic\metre}$ is the gas density at equilibrium, and $V_0$ is the equilibrium bubble volume.
The primary acoustic radiation force, exerted on the bubble by the ultrasound pressure field \(p_{\mathrm{d}}(s, t)\), is expressed as:
\begin{equation}
{F}_{\mathrm{US}}(t) =  -V \frac{\partial p_{\mathrm{d}}}{\partial s},
\end{equation}
where \(V\) is the bubble volume.
This expression holds when the acoustic pressure varies only slightly over distances comparable to the bubble size \cite{Prosperetti1982BubbleResults}.
The added mass force accounts for the movement of surrounding fluid carried along with the bubble and is given by \cite{Takemura2004TheNumber}:
\begin{equation}
{F}_{\mathrm{AM}}(t) = -\frac{1}{2} \rho_{\mathrm{l}} \left(\dot{V}\dot{s} + V\ddot{{s}} \right).
\end{equation}
This formulation remains valid for all Reynolds numbers $\text{Re} = 2 \rho_{\mathrm{l}} R \dot{s}/ \mu_{\mathrm{l}}$ and irrespective of the boundary condition (no-slip or free-slip) at the bubble surface \cite{Magnaudet2000TheFlows}.
The viscous drag force, ${F}_{\mathrm{D}}(t)$, acting on the translating bubble consists of a quasi-steady component ${F}_{\mathrm{QS}}(t)$ and a history-dependent term ${F}_{\mathrm{H}}(t)$.
For finite Reynolds numbers, the quasi-steady drag force is given by:
\begin{equation}
{F}_{\mathrm{QS}}(t) = - \frac{1}{2} \rho_{\mathrm{l}} \pi R^2 \dot{{s}}^2 C_{\mathrm{D}}(\textrm{Re}),
\end{equation}
where the drag coefficient is approximated as \cite{White1991ViscousFlow}:
\begin{equation}
C_{\mathrm{D}}(\textrm{Re}) = \frac{24}{\mathrm{Re}} + \frac{6}{1+ \sqrt{\mathrm{ Re}}} +0.4.
\end{equation}
Using the finite-Reynolds correction for the drag coefficient is essential, as the bubble can transiently reach $\mathrm{Re}\sim\mathcal{O}(10)$, where simple Stokes drag ($C_{\mathrm{D}}=24/\mathrm{Re}$) would significantly overestimate the displacement.
The history-dependent drag arises due to the unsteady diffusion of vorticity from the bubble surface. 
In the zero-Reynolds-number limit, for a no-slip surface condition (such as a surfactant-covered interface), it is given by the analytical expression \cite{Takemura2004TheNumber}:
\begin{equation}\label{eq:0Re}
{F}_{\mathrm{H}}(t) = - 6 \pi \mu_{\mathrm{l}}\int_{-\infty}^t \frac{1}{\sqrt{\pi \nu_{\mathrm{l}}\int_{\tau}^t R^{-2}dt'}}\frac{d(R \dot{{s}})}{d\tau}d\tau,
\end{equation}
where \(\nu_{\mathrm{l}}\) is the fluid kinematic viscosity.
For finite Reynolds numbers, this force can only be described using semi-empirical expressions. 
The most refined formulation available to date is given by \cite{Takemura2004TheNumber}:
\begin{equation}\label{eq:fRe}
{F}_{\mathrm{H}}(t) = - 6 \pi \mu_{\mathrm{l}}\int_{-\infty}^t \left\{\left[\pi \nu_{\mathrm{l}} \int_{\tau}^{t} R^{-2}dt'\right]^{1/2\alpha} +\frac{1}{G}\left[\frac{\pi}{16}\left(\frac{\text{Re}}{f_{\mathrm{H}}(\text{Re})}\right)^3\left(\nu_{\mathrm{l}}\int_{\tau}^tR^{-2}dt'\right)^2\right]^{1/\alpha}\right\}^{-\alpha}\frac{d(R \dot{{s}})}{d\tau}d\tau,
\end{equation}
where $\alpha =2.5$ and $f_{\mathrm{H}}(\text{Re})=0.75+0.126\text{Re}$.
The function $G$ is defined as:
\begin{equation}
G = 1 + \frac{\beta}{1+ \displaystyle\frac{(\text{Dc}/\text{Ac})^{\delta}}{\epsilon(1+(\text{Dc}/\text{Ac})^{\delta-1})}} (\text{Ac})^{1/2},
\end{equation}
where $\beta = 22$, $\delta =0.25$ and $\epsilon=0.07$.
The acceleration parameter $\text{Ac}$ is given by:
\begin{equation}
\text{Ac} = \frac{2}{\dot{s}^2}\left| \frac{d(R \dot{{s}})}{d\tau}\right|.
\end{equation}
The dimensionless time rate of change of acceleration, $\text{Dc}$, is expressed as:
\begin{equation}
\text{Dc} = \frac{4}{R\left|\dot{s}^3\right|}\left| \frac{d}{d\tau} \left(R^2 \frac{d(R \dot{{s}})}{d\tau}\right) \right|.
\end{equation}
The history integral exhibits a singularity at its upper limit.
Nevertheless, it can still be evaluated using the numerical method proposed by Chung \cite{Chung1982TheDroplet,Kim1998OnNumbers}.
We note that the no-slip condition at surfactant-covered interfaces arises from Marangoni stresses, which are not accounted for in the radial dynamics equation.
Furthermore, we do not consider variations in the no-slip condition due to changes in lipid coverage during the bubble radial excursion.
These approximations enable a more manageable formulation of the problem.
To solve the radial and translational equations of motion, we can exploit the fact that bubble displacement has a negligible influence on the radial dynamics.
This allows us to decouple the problem: we can first solve the radial motion equation and then, using that solution, we can solve the translational motion equation rather than treating the two as fully coupled.
The bubble is initially at rest, so the initial conditions are: $R(0) = R_0, \ \dot{R}(0)=0, \ p_{\rm g}(0) = P_{\infty} + 2\sigma_0/R_0, \ s(0) = 0, \ \dot{s}(0) = 0$.

\begin{figure}[ht]
    \includegraphics[width = 0.5\columnwidth]{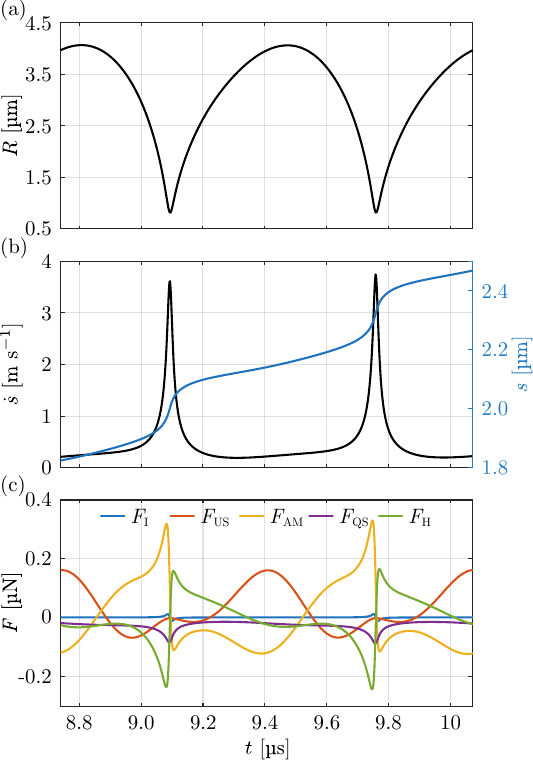}
    \centering
     \caption{Representative simulation of the radial and translational dynamics, and of the corresponding force balance in the direction of motion, for a free bubble with equilibrium radius \( R_0 = \SI{2.44}{\micro\meter} \) driven by an ultrasound pulse at $f_{\rm d} = \SI{1.5}{\mega\hertz}$ and pressure amplitude \( p_{\mathrm{a}} = \SI{100}{\kilo\pascal} \).
     The time histories span two ultrasound cycles.
     (a) Time evolution of the bubble radius.
     (b) Time evolution of the bubble translation speed and displacement.
     (c) Forces acting on the bubble in the direction of translational motion: $F_{\rm I}$ inertial force, $F_{\rm US}$ primary acoustic radiation force, $F_{\rm AM}$ added mass force, $F_{\rm QS}$ quasi-steady drag force, $F_{\rm H}$ history drag force.
     }\label{fig:Figure2}
\end{figure}
Figure~\ref{fig:Figure2} shows a representative example of the simulated time histories—spanning two ultrasound periods—of (a) the radius of an ultrasound-driven microbubble, (b) its translational velocity and displacement, and (c) the forces acting on the bubble along the direction of motion.
The translational velocity increases sharply during the compression phases and reaches its maximum when the bubble radius is minimum.
Interestingly, this velocity peak is not driven by the primary acoustic force, which is nearly zero at minimum radius, but instead by the added-mass response of the surrounding fluid, which impulsively drives the bubble as its volume decreases rapidly.
Over a full acoustic cycle, the acoustic-radiation and added-mass contributions to the net displacement are comparable; however, the added-mass contribution is strongly localised around the instants of minimum radius.
The quasi-steady drag consistently opposes the motion and peaks when the translational velocity is largest, whereas the history-dependent drag counteracts the bubble accelerations and decelerations.
Finally, the bubble inertial force is negligible relative to the other contributions and could be omitted; nevertheless, it is retained in the formulation and in the simulations below for completeness.

\section{Results and discussion}

\begin{figure}[H]
    \includegraphics[width = \columnwidth]{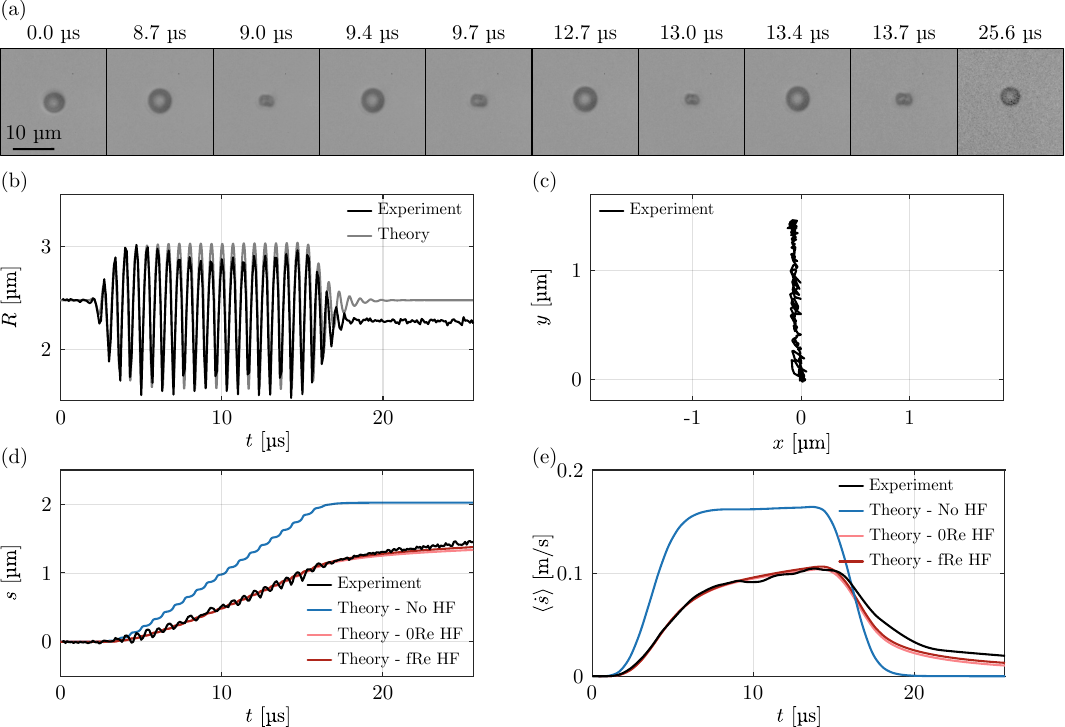}
     \caption{
     Radial and translational motion of a bubble (equilibrium radius \( R_0 = \SI{2.48}{\micro\meter} \)) in free space, driven by a 20-cycle ultrasound pulse at a frequency $f_{\rm d} = \SI{1.5}{\mega\hertz}$ with a pressure amplitude of \( p_{\mathrm{a}} = \SI{50}{\kilo\pascal} \).
     (a) Image sequence showing the bubble response to the ultrasound pulse, extracted from Supplementary Movie 1.
     The bubble exhibits buckling during compression phases.
     No shape mode oscillations occur.
     (b) Comparison between the experimentally measured bubble radius over time and the theoretical prediction (Eq.~\(\ref{eq:RP}\)).
     (c) Trajectory of the bubble translational motion in the horizontal \(xy\)-plane.  
     (d) Comparison between the experimentally measured bubble displacement (measured along the straight line from its initial to final position) over time and three theoretical models: one without the history force (No~HF), one incorporating the zero-Reynolds-number approximation of the history force (0Re~HF, Eq.~\ref{eq:0Re}), and one including its finite-Reynolds-number extension (fRe~HF, Eq.~\ref{eq:fRe}).
     (e) Comparison between the experimentally measured time evolution of the bubble averaged displacement velocity and the predictions from the same three models: No~HF, 0Re~HF, and fRe~HF.
     The uncertainty in the experimental measurements corresponds to half the pixel size (\SI{80}{\nano\meter}).
     }\label{fig:Figure3}
\end{figure}

Figure \ref{fig:Figure3}(a-e) and Supplementary Movie 1 illustrate the radial and translational dynamics of a bubble with an equilibrium radius of \( R_0 = \SI{2.48}{\micro\meter} \) in free space, subjected to a 20-cycle ultrasound pulse at \( \SI{1.5}{\mega\hertz} \) and a pressure of \( p_{\mathrm{a}} = \SI{50}{\kilo\pascal} \).
This bubble size approximately corresponds to the resonant size for the applied driving frequency.
When the ultrasound pulse is active, the measured averaged Reynolds number associated with the bubble translation is \( \text{Re} = 0.36 \), while the Reynolds number associated with the bubble oscillation is \( \mathcal{U}\text{Re} = 17.7 \), where $\mathcal{U} = \dot{R} / \dot{s}$.
As shown in Supplementary Movie 1, with selected frames displayed in Fig.~\ref{fig:Figure3}(a), the bubble exhibits purely spherical oscillations under this pressure driving, with a clear shell buckling instability that induces a brief departure from sphericity during the compression phase.
The normalised radial expansion of the bubble is $\Delta R/R_0 = 0.21$, corresponding to a normalised volumetric expansion of $\Delta V/V_0 = 0.75$.
The radial expansion $\Delta R/R_0$ is determined by calculating the difference between the maximum bubble radius—averaged over oscillation cycles once steady state is reached—and the equilibrium radius.
No non-spherical oscillations (shape modes) are observed, as the radial expansion remains insufficient to trigger the Faraday instability.
As demonstrated in previous studies \cite{Cattaneo2023ShellMicrobubbles,Spiekhout2024Aremono-acoustic}, the radial dynamics is strongly influenced by the shell rheological properties, particularly its viscosity. 
Consequently, the shell viscosity used in our simulations is fitted against the experimental radial dynamics, resulting in a value of $\kappa_{\mathrm{s}} = \SI {5e-9} {\kilo\gram\per\second}$.
The shell elasticity and initial surface tension are set for all simulations to $ E_{\mathrm{s}} = \SI{0.6}{\newton\per\meter}$ and $\sigma_0 = \SI{0}{\newton\per\meter}$, respectively—values that provided the best overall fit across different bubble sizes in our previous study on the rheological characterisation of phospholipid shells.
Figure~\ref{fig:Figure3}(b) compares the numerical predictions (Eq.~\ref{eq:RP}), derived from the fitted shell parameters, with experimental results for radial dynamics.
A distinct decrease in bubble radius is observed at the end of the ultrasound pulse, which can be attributed to gas efflux during compression phases when the gas core pressure increases, as well as lipid shedding occurring during bubble oscillation.
These mechanisms are not accounted for in the numerical model.
The trajectory of the bubble centroid follows a straight line aligned with the direction of the ultrasound, as shown in Fig.~\ref{fig:Figure3}(c).
The displacement of the bubble over time, denoted as \( s(t) \), is presented in Fig.~\ref{fig:Figure3}(d).
The displacement is measured along the straight line connecting the bubble initial and final positions.
During the ultrasound pulse, the displacement exhibits an approximately linear progression over time, followed by a slower drift after the pulse ceases.
This continued motion is due to the history force, which resists deceleration and thus prolongs the bubble movement, ultimately contributing to about 30\% of the total displacement.
Additionally, during the pulse-on phase, a subtle back-and-forth motion of the bubble is observed, oscillating at the same frequency as the ultrasound.
Theoretical predictions for bubble displacement, obtained from Eq.~(\ref{eq:forcebalance}) using the same shell parameters fitted to radial dynamics, are compared with experimental measurements.
The analysis considers three distinct models for the drag force.
The first model accounts only for the quasi-steady contribution, excluding the history force.
The second incorporates the history force in its form valid for zero Reynolds numbers (0Re), while the third includes the history force formulation applicable to finite Reynolds numbers (fRe).  
The model without the history force significantly overestimates the bubble displacement, by 75\% when the final drift motion is not considered and by 40\% when it is included.
Additionally, this model predicts that the bubble comes to an abrupt stop once the ultrasound driving ends.  
In contrast, the model incorporating the 0Re-history force exhibits strong agreement with experimental data across all phases of the displacement. 
Similarly, the fRe-history force model produces nearly identical results, confirming that, at the present Reynolds numbers, the bubble dynamics can be effectively described using the zero-Reynolds-number theory.  
However, none of the models successfully reproduce the weak back-and-forth bubble motion observed in the experiment when the ultrasound driving is on.
This motion is likely due to fluid particle velocity induced by the travelling ultrasound wave ($u_{\rm p} = p_{\rm d}/\rho_{\rm l}c_{\rm l}$), which is not included in the model.
The time evolution of the bubble averaged translational velocity, $\langle\dot{s}\rangle$, is shown in Fig.~\ref{fig:Figure3}(e).
The velocity stabilises to a nearly steady state within ten ultrasound cycles, reaching a value roughly a hundred times higher than the blood flow speed in capillaries.
This indicates that the bubble can easily overcome capillary flow and effectively target therapeutic sites.
The theoretical model without the history force predicts a higher steady-state velocity, a faster transition to this velocity, and a more abrupt stop once the ultrasound pulse ends, compared to the models that include the history force. 
In contrast, these latter models accurately capture the initial transient acceleration, as well as the weaker acceleration during the pulse-on phase. 
They slightly underestimate the bubble velocity during the final drift but overall provide a much closer match to the experimental observations.

\begin{figure}[h]
    \includegraphics[width = \columnwidth]{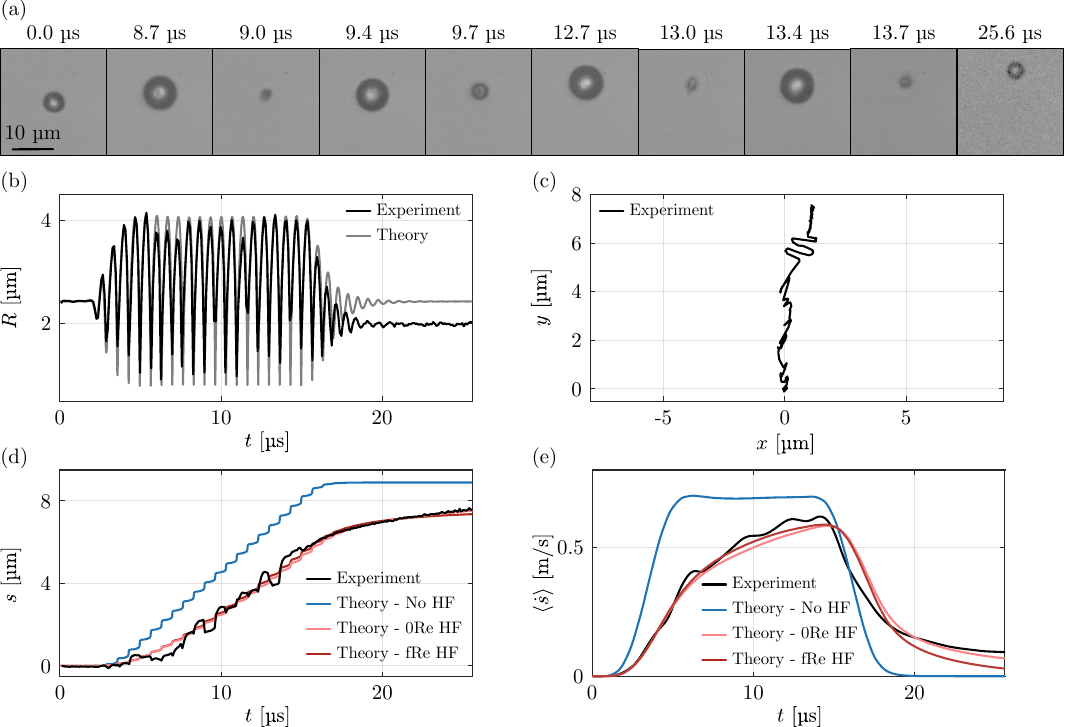}
     \caption{Radial and translational motion of a bubble (equilibrium radius \( R_0 = \SI{2.44}{\micro\meter} \)) in free space, driven by a 20-cycle ultrasound pulse at a frequency $f_{\rm d} = \SI{1.5}{\mega\hertz}$ with a pressure amplitude of \( p_{\mathrm{a}} = \SI{100}{\kilo\pascal} \).
     (a) Image sequence showing the bubble response to the ultrasound pulse, extracted from Supplementary Movie 2.
     The bubble displays a distinct alternating rigid-body motion at half the ultrasound frequency, overlaid on its overall translational movement.
     This motion results from the excitation of the shape mode with wavenumber \(l=1\), induced by the Faraday instability.
     (b) Comparison between the experimentally measured bubble radius over time and the theoretical prediction (Eq.~\(\ref{eq:RP}\)).
     (c) Trajectory of the bubble translational motion in the horizontal \(xy\)-plane.
     (d) Comparison between the experimentally measured bubble displacement (measured along the straight line from its initial to final position) over time and three theoretical models: one without the history force (No~HF), one incorporating the zero-Reynolds-number approximation of the history force (0Re~HF, Eq.~\ref{eq:0Re}), and one including its finite-Reynolds-number extension (fRe~HF, Eq.~\ref{eq:fRe}).
     (e) Comparison between the experimentally measured time evolution of the bubble averaged displacement velocity and the predictions from the same three models: No~HF, 0Re~HF, and fRe~HF.
     The uncertainty in the experimental measurements corresponds to half the pixel size (\SI{80}{\nano\meter}).
     \label{fig:Figure4}
     }
\end{figure}
A similar-sized bubble, with an initial radius of \(R_0 = \SI{2.44}{\micro\meter}\), is tested at a higher ultrasound pressure of \(p_{\mathrm{a}} = \SI{100}{\kilo\pascal}\).
Its radial and translational dynamics are presented in Fig.~\ref{fig:Figure4}(a–e) and Supplementary Movie 2.
The normalised radial expansion of the bubble is $\Delta R/R_0 = 0.65$, corresponding to a normalised volumetric expansion of $\Delta V/V_0 = 3.42$.
As shown in Supplementary Movie 2, with selected frames displayed in Fig.~\ref{fig:Figure4}(a), the bubble exhibits a pronounced alternating rigid-body motion at half the ultrasound frequency, superimposed on its overall translational motion.
This behaviour arises from the excitation of the non-spherical mode (also termed shape mode) with wavenumber \(l=1\), triggered by the Faraday instability when the radial excursion—and consequently, the acceleration—of the bubble interface exceeds a critical threshold.
The characteristic half-harmonic response is a key signature of this instability.
For a fixed driving frequency, the wavenumber of the excited shape mode increases with bubble size.
Under the applied driving conditions, bubbles near the resonant size exhibit the \(l=1\) mode.
For a more detailed discussion of this interfacial instability, please refer to our previous works \cite{Cattaneo2025CyclicDelivery,Cattaneo2025ShapeBubbles,Cattaneo2025FaradayJetting}.
In this example, the average Reynolds number associated with the bubble translational motion during the ultrasound pulse is \( \text{Re} = 1.96 \), while the Reynolds number corresponding to its oscillatory motion is \( \mathcal{U}\text{Re} = 43.4 \).
Despite the presence of shape modes, the radial motion model still shows good agreement with experimental results as shown in Fig.~\ref{fig:Figure4}(b).
The apparent discrepancy during the compression phases is attributed to the limited temporal resolution of the imaging system, which may miss the very sharp minima characteristic of high-pressure cuspidal compressions, leading to an underestimation of the compression amplitude.
The shell viscosity obtained from the fitting procedure is $\kappa_{\mathrm{s}} = \SI {2e-9} {\kilo\gram\per\second}$, which remain well within the range reported in our previous shell rheological assessment \cite{Cattaneo2023ShellMicrobubbles}.
Meanwhile, as illustrated in Fig.~\ref{fig:Figure4}(c), the bubble trajectory is more erratic due to the zig-zag motion at half the ultrasound frequency induced by the shape mode.
Figure~\ref{fig:Figure4}(d) demonstrates once again that the model for bubble translation, when excluding the history force, overestimates the bubble displacement.
However, in this case, the discrepancy is reduced to 40\% without considering the final drift and 17\% when it is included.  
Notably, even in the presence of shape modes, the models incorporating the history force accurately capture the bubble displacement.
This suggests that the net effect of the alternating rigid body motion on bubble translation is approximately negligible.  
Moreover, the difference between the two models that include the history force remains marginal.
This extends the validity of the zero-Reynolds-number approximation even at the present finite Reynolds number, which can be considered the upper limit for bubble-delivery applications.
Figure~\ref{fig:Figure4}(e) illustrates the time evolution of the averaged bubble translational velocity, showing that the difference in steady-state velocity between theoretical models with and without the history force decreases as the number of cycles increases. 
This suggests that for longer ultrasound pulses, the difference in displacement between the actual motion and the prediction without the history force becomes minor, supporting the use of simplified models as a first approximation for these Reynolds numbers.
Previous studies have suggested that the model incorporating the history force is valid only for \( Re < 1 \) and \( URe < 14 \) \cite{Magnaudet1998TheRadius}.
However, our findings indicate that it remains valid for higher Reynolds numbers.
This is not surprising, as previous investigations focused on markedly different conditions—rapidly shrinking, rising bubbles in a quiescent liquid—unlike the acoustically driven microbubbles examined in our work.
Moreover, the validity of the 0Re-history force translational model at Reynolds numbers slightly above unity has also been noted by Acconcia \textit{et al.} \cite{Acconcia2018TranslationalPulses}.

\begin{figure}[t]
    \includegraphics[width = \columnwidth]{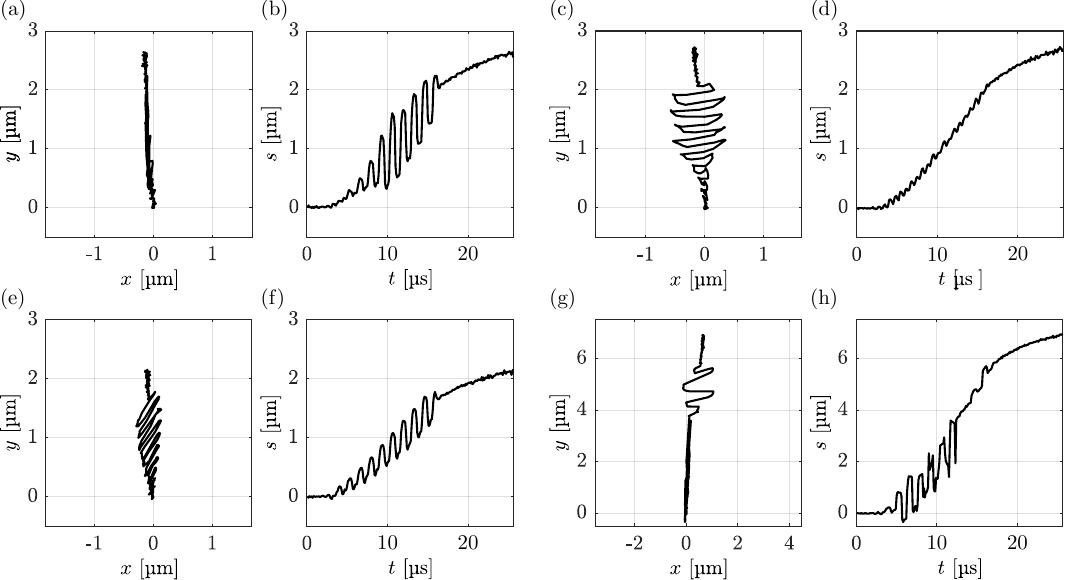}
     \caption{Trajectories and displacements over time for four different cases of ultrasound-driven bubbles exhibiting the \( l=1 \) shape mode caused by the Faraday instability, illustrating the varying directions of the associated rigid-body motion:
     (a-b) Motion aligned with the ultrasound propagation direction.
     (c-d) Motion perpendicular to the ultrasound propagation direction.
     (e-f) Motion oriented obliquely relative to the ultrasound propagation direction.
     (g-h) Motion that transitions during the ultrasound pulse—from parallel to perpendicular relative to the propagation direction.
     The amplitude of the rigid-body motion increases with ultrasound cycles until saturation, a characteristic feature of a parametric instability, such as the Faraday instability.
     The uncertainty in the experimental measurements corresponds to half the pixel size (\SI{80}{\nano\meter}).
     }
     \label{fig:Figure5}
\end{figure}
The appearance of shape modes has not been reported in previous studies on the translation dynamics of microbubbles, likely due to the absence of time-resolved radial dynamics or to the use of streak cameras.
The direction of the alternating rigid-body motion induced by the shape mode \( l=1 \), which is particularly relevant as it occurs at resonant sizes for the employed driving frequency, is largely unpredictable, since the system is nearly perfectly spherically symmetric, and depends only weakly on the ultrasound propagation direction.
Figure~\ref{fig:Figure5} and Supplementary Movies 3-6 present different instances of trajectories and displacements over time of bubbles exhibiting the \( l=1 \) shape mode.
The direction of motion can align parallel to the ultrasound direction (Fig.~\ref{fig:Figure5}(a-b) and corresponding Supplementary Movie 3), be perpendicular to it (Fig.~\ref{fig:Figure5}(c-d) and corresponding Supplementary Movie 4), or take any intermediate angle (Fig.~\ref{fig:Figure5}(e-f) and corresponding Supplementary Movie 5).
Generally, this direction remains constant throughout the pulse duration; however, when the shape mode is particularly intense, it can shift by as much as \ang{90} (Fig.~\ref{fig:Figure5}(g-h) and corresponding Supplementary Movie 6).
The magnitude of the alternating motion increases with successive ultrasound cycles until it stabilises at a steady-state amplitude of approximately $\SI{0.5}{\micro\meter}$.
This saturation behaviour is characteristic of parametric instabilities, such as the Faraday instability.
The minimal influence of shape modes on the overall displacement of the bubble is evident in cases where the alternating rigid-body motion is misaligned with the ultrasound direction (Fig.~\ref{fig:Figure5}(c-f)).
Even after multiple shape mode cycles, the overall displacement direction remains unchanged, consistently aligning with the direction of ultrasound propagation.

\begin{figure}[t]
    \includegraphics[width = 0.5\columnwidth]{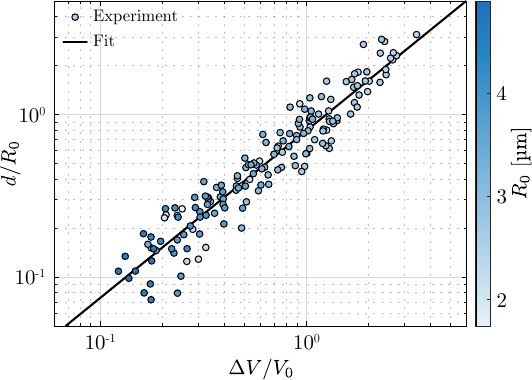}
     \caption{Normalised bubble displacement, ${d}/{R_0}$, as a function of normalised bubble volumetric expansion, ${\Delta V}/{V_0}$, on a logarithmic scale for $n = 162$ data points, obtained from bubbles spanning a range of equilibrium radii $R_0$, each driven by a 20-cycle, $f_{\rm d} = \SI{1.5}{\mega\hertz}$ ultrasound pulse at varying pressure amplitudes.
     The least-squares fit line has a slope of 1.03, indicating a linear relationship between normalised bubble displacement and normalised volumetric expansion.     
     \label{fig:Figure6}
     }
\end{figure}
We now seek a scaling law for bubble displacement.
We propose that the normalised bubble displacement is proportional to a power \(\alpha\) of the normalised bubble volumetric expansion, i.e. ${d}/{R_0} \propto \left({\Delta V}/{V_0}\right)^\alpha$.
To determine \(\alpha\), we perform a least-squares fit of \(d/R_0\) against \(\Delta V/V_0\) using $n=162$ data points from bubbles spanning a range of equilibrium radii, each driven by a 20-cycle, 1.5-MHz ultrasound pulse at varying pressure amplitudes.
Figure~\ref{fig:Figure6} displays the measured \(d/R_0\) values as a function of \(\Delta V/V_0\) on a logarithmic scale, along  with the linear regression line.
The fit yields \(\alpha = 1.03\), indicating that the normalised displacement scales linearly with the normalised volumetric expansion:
\begin{equation}
\frac{d}{R_0} \sim \frac{\Delta V}{V_0}.
\end{equation}
This linear scaling is consistent with the structure of the translational model: the forcing terms depend directly on the time-varying bubble volume, while added-mass and drag contributions remain approximately linear in the displacement dynamics over the explored parameter range.
This result is striking in its simplicity and provides a practical tool for predicting the bubble displacement from its radial excursion, or vice versa.
Naturally, the absolute magnitude of the normalised displacement depends on the number of ultrasound cycles used.
If the same volume expansion could be maintained, doubling the pulse length would therefore double the distance travelled by the bubble.
In practice, however, sustaining constant volume expansion over longer pulses is nontrivial. As the bubble dissolves, its radius decreases with time, which can detune it from resonance. The oscillation amplitude may then decline, reducing the volume expansion and yielding a smaller displacement than would be predicted by a simple linear scaling.
By dividing the normalised displacement by the number of cycles $n$, we obtain a normalised average velocity expressed as a function of the normalised volumetric expansion:
\begin{equation}
\tilde{v} = \frac{d}{R_0} \frac{1}{n} \approx 0.04 \, \frac{\Delta V}{V_0}.
\end{equation}
The fit in Fig.~\ref{fig:Figure6} spans multiple acoustic pressures and inherently includes bubble-to-bubble variations in shell rheology.
It is expected to be largely independent of shell lipid formulation: for a given volumetric expansion history, the hydrodynamic forces governing translation are essentially unchanged, so the net displacement should be similar across lipid-coated agents.
This may not hold for bare bubbles (free slip boundary condition) or for coatings with markedly rough/structured surfaces, which could modify the effective drag.
All measurements are performed at a single ultrasound frequency, so generalisation to other frequencies requires additional validation and represents an important direction for future work.
Shape modes of varying degree $l$ (with $l$ increasing with bubble size) are present in our dataset.
However, we find no measurable dependence of the net translational displacement on these modes: their effect is purely oscillatory and averages to zero over a full shape mode cycle.

\begin{figure}[t]
\includegraphics[width = 0.5\columnwidth]{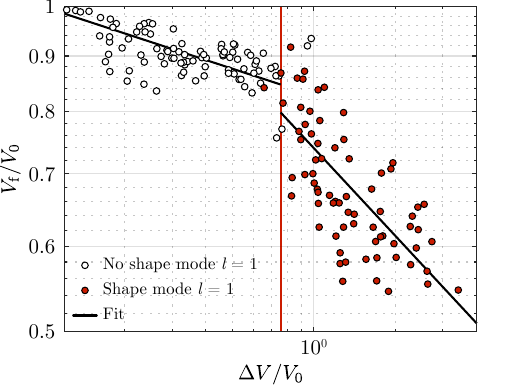}
     \caption{Ratio of the bubble volume after the ultrasound pulse to its equilibrium volume, \(V_{\mathrm{f}}/V_0\), as a function of normalised bubble volumetric expansion, \(\Delta V/V_0\), on a logarithmic scale based on $n=162$ data points from bubbles with varying equilibrium radii, \(R_0\)), all subjected to a 20-cycle ultrasound pulse at a frequency $f_{\rm d} = \SI{1.5}{\mega\hertz}$.
     Solid lines denote the least-squares fits.
     The bubble dissolution rate triples when \(\Delta V / V_0\) exceeds approximately 0.75.
     Beyond this threshold, bubbles begin to exhibit the shape mode \(l=1\), and in the most forceful cases, the associated cyclic jetting (for details, see Ref. \cite{Cattaneo2025CyclicDelivery}).
     The more pronounced lipid shedding presumably associated with these dynamics likely explains the increased dissolution rate.   
     \label{fig:Figure7}
     }
\end{figure}
Finally, we explore the trade-off between rapid bubble transport and bubble stability. 
Figure~\ref{fig:Figure7} plots the ratio of the final bubble volume as measured after the ultrasound pulse to its equilibrium volume, \(V_{\mathrm{f}}/V_0\), against the normalised volumetric expansion, \(\Delta V/V_0\), which, as previously shown, is linearly correlated with bubble displacement.
All measurements are obtained using a 20-cycle ultrasound pulse.
As a first-order approximation, we expect the volume loss due to dissolution to increase approximately linearly with pulse length. 
However, dissolution of coated bubbles under ultrasound is a complex phenomenon which exhibits substantial variability, as evidenced by the scatter in the data in Fig.~\ref{fig:Figure7}.
As a result, stronger conclusions require further experiments.
The data indicate that for \(\Delta V/V_0 \lesssim 0.75\), the bubbles dissolve following the scaling law:
\begin{equation}
\frac{V_{\mathrm{f}}}{V_0} \sim \left(\frac{\Delta V}{V_0}\right)^{0.08}.
\end{equation}
However, above this threshold the dissolution rate more than triples, with the data fitting:
\begin{equation}
\frac{V_{\mathrm{f}}}{V_0} \sim \left(\frac{\Delta V}{V_0}\right)^{0.27}.
\end{equation}
The marked increase in dissolution correlates with the appearance on the bubble surface of a shape mode with wavenumber \(l=1\).
This specific mode emerges because the bubbles experiencing the largest expansions are near their resonant size, and for resonant bubbles, the \(l=1\) shape mode dominates among all possible modes.
When this mode manifests, it is accompanied by a significant asymmetric collapse and, in some instances, cyclic jetting, as we have previously reported \cite{Cattaneo2025CyclicDelivery}. 
These events likely induce considerable lipid shedding, strongly accelerating bubble dissolution.
This interpretation is supported by the observation that the threshold identified here is comparable to the lipid-shedding threshold reported by Luan \textit{et al.}\cite{Luan2014}.
The increased scatter in this higher dissolution-rate regime likely stems from the inherently nonlinear process of shell shedding.
Small variations in the timing and extent of shedding—across cycles and between individual bubbles—can produce large differences in the measured dissolution rates.
Since shell shedding is also likely influenced by the shell lipid composition, the degree of dissolution observed at a given volumetric expansion may vary across different shell lipid formulations.
Based on these findings, an optimal bubble transport strategy should avoid exceeding the volumetric expansion threshold that triggers the \(l=1\) shape mode, since crossing this limit greatly reduces bubble stability.
Instead, a small volumetric expansion is preferable, given that high transport speeds are unnecessary in capillaries—where most therapeutic action occurs—due to their narrow diameter and the relatively slow blood flow of approximately \SI{0.001}{\meter\per\second} \cite{Sidebotham2007ChapterPathophysiology}.
Moreover, short ultrasound pulses are advisable since bubbles reach steady-state velocity within fewer than ten ultrasound cycles, allowing repeated pulses to leverage the substantial drift observed during the off phases of ultrasound driving.

\section{Conclusions}

This study investigates the translational dynamics of phospholipid-coated microbubbles driven by the primary acoustic radiation force induced by an ultrasound pulse.
Using optical tweezers to manipulate bubbles in three-dimensional space, we were able to interrogate bubble dynamics without interference from nearby walls.
Ultra-high-speed imaging allowed us to time-resolve both radial and translational dynamics. 
Theoretical models for radial dynamics, based on the Rayleigh–Plesset equation, and for translational dynamics, based on the balance of forces acting on the bubble, showed good agreement with experimental observations for the bubble sizes studied.
For the Reynolds numbers considered (up to $\text{Re}\approx2$), incorporating the history drag force was necessary to achieve accurate predictions.
No significant difference was found between the zero-Reynolds-number and finite-Reynolds-number formulations of the history force within the $\text{Re}$ range investigated.
Non-spherical modes, triggered by the Faraday instability, emerged beyond a certain volumetric bubble expansion.
However, their overall impact on translational motion was negligible. 
Notably, we found that the bubble displacement, normalised by its equilibrium radius, ${d}/{R_0}$, scaled linearly with the bubble volumetric expansion, normalised by its equilibrium volume, ${\Delta V}/{V_0}$.
This scaling law provides a simple and practical way to connect radial and translational dynamics.
However, beyond a threshold of ${\Delta V}/{V_0}\approx 0.75$, the bubble dissolution rate increased significantly due to the onset of shape modes and cyclic jetting, likely leading to lipid shedding.
Based on these findings, we recommend using mild, short, and repeated ultrasound pulses as an optimal transport strategy for coated microbubbles to balance transport speed with bubble stability.
The main limitation of this study is that it uses water as the surrounding medium, while therapeutic microbubbles typically operate in blood. Blood is a multicomponent suspension of red blood cells in plasma, and because microbubble sizes are often comparable to those of red blood cells, direct bubble–cell interactions can significantly affect the effective resistance to motion. Additionally, plasma is more viscous than water, leading to greater viscous dissipation and damping of both radial oscillations and translation. As a result, for the same volumetric expansion amplitude, bubble displacement in blood would generally be lower due to the increased resistance.
Finally, in biologically relevant environments, coated microbubbles may also experience confinement, strong background flows in large vessels, shear flow, variations in ambient pressure, and interactions with other microbubbles, all of which can further modulate their oscillatory dynamics and net translation.

\section{Acknowledgments} We thank ETH Zurich for financial support.

\section{Data availability} The data that support the findings of this article are openly available \cite{Cattaneo2025SourceUltrasound}.

\bibliography{main}

\end{document}